\documentclass[aps,prl,epsfigure,twocolumn,superscriptaddress]{revtex4-1}
\usepackage{dcolumn} 
\usepackage{bm} 
\usepackage{graphicx}
\usepackage{amsmath}    
\usepackage{latexsym}
\usepackage{amsfonts}   
\usepackage{amssymb}
\usepackage{array}      
\usepackage{epsfig}
\usepackage{txfonts}
\usepackage{color}
\usepackage[colorlinks=true,linkcolor=blue,urlcolor=blue,citecolor=blue]{hyperref}
\usepackage{hyperref}
\usepackage{enumerate}
\usepackage{bbold}
\usepackage{cleveref}
\usepackage{appendix}

\newcommand{\xis}{\xi_h}
\newcommand{\nus}{\nu_h}
\newcommand{\ke}{\kappa_{\rm ex}}

\newcommand{\ntu}{Division of Physics and Applied Physics, School of Physical 
	and Mathematical Sciences, Nanyang Technological University, Singapore}
\newcommand{\majulab}{MajuLab, CNRS-UCA-SU-NUS-NTU International Joint Research
	Unit, Singapore}
\newcommand{\cnr}{CNR--SPIN, Dipartimento di Scienze Fisiche, Universit\`a di 
	Napoli Federico II, I-80126, Napoli, Italy}

\newcommand{\cnrs}{Laboratoire Charles Coulomb (L2C), Université de Montpellier, CNRS, 34095 Montpellier, France}
\begin{document}
\title{Hidden Order Beyond Hyperuniformity in Critical Absorbing States}
\author{Yuanjian Zheng} 	
\affiliation{\ntu}
\author{Anshul D. S. Parmar}
\affiliation{\cnrs}
\author{Massimo Pica Ciamarra} 
\affiliation{\ntu}
\affiliation{\majulab}
\affiliation{\cnr}

\begin{abstract}     
Disordered hyperuniformity is a description of hidden correlations in point distributions revealed by an anomalous suppression in fluctuations of local density at various coarse-graining length scales. 
In the absorbing phase of models exhibiting an active-absorbing state transition, this suppression extends up to a hyperuniform length scale that diverges  at the critical point.
Here, we demonstrate the existence of additional many-body correlations beyond hyperuniformity.
These correlations are hidden in the higher moments of the probability distribution of the local density, and extend up to a longer length scale with a faster divergence than the hyperuniform length on approaching the critical point.
Our results suggest that a hidden order beyond hyperuniformity may generically be present in complex disordered systems.
\end{abstract}	
\maketitle

The behaviour of long-wavelength density fluctuations and their anomalous suppression - hyperuniformity \cite{Torquato2018, Torquato2016}, have been a subject of recent interest in the study of disordered systems for they provide an avenue to probe long-range order in problems that do not possess translational or bond-orientational symmetry \cite{KlattTorquato2019, Torquato2018b, Hexner2018, HexnerLevine2015, HexnerLevine2017, ChiecoDurian2018,Mitra2020}. 
At the same time, hyperuniformity is also emergent in a diverse variety of naturally occurring or model systems, that range from granular or colloidal materials \cite{Torquato2018b, Hexner2018, WilkenChaikin2020, WangPaulsen2018, TjhungBerthier2015, ChremosDouglas2018} to soft biological tissues \cite{JiaoTorquato2014, ZhengPicaCiamarra2020, LiBi2018}. This has led to speculations on its universality \cite{KlattTorquato2019, HexnerLevine2015} and the need for greater understanding of its causal role in the organization and structure of complex systems. 

For a given configuration of points ${\vec{r}_i} $ in $d$-dimensional space with the global number density $\rho$, the {\it local} density $\rho_R \equiv \frac{1}{R^d}\sum_{\vec{r}_j \in \Omega} \delta(\vec{r}_i- \vec{r}_j)$,   
defined over a subspace region $\Omega $ of some finite length scale $R$ is a coarse-grained variable characterized by a discrete probability distribution $P(\rho_R)$. 
In the scenario where $\vec{r}_i$ is generated randomly by an underlying Poisson point process, $\rho_R$ of disconnected regions in real space are uncorrelated such that $P(\rho_R)$ is constrained by the central limit theorem (CLT) and its variance scales as $\sigma^2(R) \sim R^{-d}$ in the limit of $R \to \infty$. 
Hyperuniformity is the characterization of density fluctuations $\sigma^2(R) \sim R^{-a} $ that are anomalously suppressed ($ a > d$) even in the thermodynamic limit due to the presence of peculiar correlations in physical density fields. 
For systems that are not (ideally) hyperuniform, $\sigma^2(R)$ is instead suppressed up to a finite length scale, $\xis$ \cite{Hexner2018, ZhengPicaCiamarra2020, ChiecoDurian2018}, which we refer to hereafter as the hyperuniform length scale.  
Hyperuniformity analyses, therefore, typically focus on characterizing pairwise correlation through $\sigma^2(R) $ and its Fourier equivalent - the structure factor $S(k)$ \cite{Torquato2016,Torquato2018}.

However, the analysis of the density field through the mere investigation of $\sigma^2(R)$ is unable to uncover the presence of additional higher-ordered correlations hidden in the higher moments~\cite{TorquatoKlatt2020} of the 
probability distribution $P(\rho_R)$.
These correlations are especially relevant at intermediate length scales where crucial information of the phase behaviour on the approach to a critical point may often be present \cite{IkedaBerthier2015, WuTeitel2015, TjhungBerthier2015, HexnerLevine2015}.

In this Letter, we show that additional many-body correlations are indeed generically hidden in the $P(\rho_R)$ of absorbing states in the Conserved Lattice Gas (CLG)~\cite{Lubeck2003, Lubeck2004, RossiVespignani2000} and the Random Organization (RO)~\cite{Tjhung2015} models.
These models undergo an active-absorbing state transition at a critical density $\rho_c$; In the high density active phase, the fraction of particles deemed  active scales as $|\rho-\rho_c|^\beta$, while dynamical correlations extend up to a length scaling as $|\rho-\rho_c|^{-\nu_\perp}$; At lower densities, these systems evolve toward an absorbing state where there are no active particles at long but finite times. 

These models have been suggested to exhibit a hyperuniformity crossover at the critical point~\cite{HexnerLevine2015}, where density fluctuations that were suppressed in the absorbing phase up to a length scale 
$\xis \propto |\rho_c-\rho|^{-\nu_\perp}$ diverge.
Here, we first demonstrate that density fluctuations in the absorbing phase are in fact only suppressed up to a length scale  $\xis \propto |\rho_c-\rho|^{-\nus}$, with $\nus < \nu_\perp$. 
Furthermore, through the investigation of the higher moments of $P(\rho_R)$ and real space analyses, we then demonstrate the existence of additional many-body correlations beyond hyperuniformity that extend up to a length scale $\xi_{\rho_R} \propto |\rho_c-\rho|^{-\nu_\perp} > \xis$.
The existence of additional correlations that are not captured by fluctuations in density fields uncovers the presence of a new form of hidden correlations beyond hyperuniformity that may prove to be generically present in complex disordered systems, such as with dynamical heterogeneities \cite{Berthier2011a} that are generically present in systems exhibiting glassy dynamics.

In the CLG model, particles are placed initially at random on an $L \times L$ square lattice with mean density $\rho \equiv N/L^2$. 
A particle is deemed active if one or more of its immediate neighbouring sites is occupied, and
active particles with at least one adjacent empty site move in parallel in each time step, randomly to an adjacent unoccupied site such that the global density $\rho$ of the system is conserved throughout its dynamics \cite{RossiVespignani2000, Lubeck2003,Lubeck2004,HexnerLevine2015}. 
This process mimics short range repulsion in conserved systems.
In the RO model~\cite{Tjhung2015}, $N$ disks of diameter $d=1$ occupy a square simulation domain of side length $L$, at volume fraction $\phi = \rho \pi (d/2)^2$ where
overlapping disks are deemed active.
At each time step, active disks move by uncorrelated displacements of random orientation and magnitude that is uniformly distributed across the range $[0,\delta]$, in which $\delta = 0.5$ is considered for this Letter.
This RO model is inspired by recent experimental investigations into colloidal suspensions under oscillatory shear~\cite{Pine2005,Corte2008}, that have recently been suggested to be hyperuniform at their active-absorbing transition~\cite{WilkenChaikin2020}.

For the CLG model, numerical results presented in the entirety of this work are for systems of linear size $L=8192$, if not otherwise stated, and $\approx 15 \times 10^6$ particles. 
These systems are up to 2 orders of magnitude larger than what has been previously considered in the literature.
For the RO model, we consider system sizes of typically $N = 8192 \times 10^3$ disks, which again is well above what has been previously considered in the literature. 
Such large systems are needed due to the strong influence of size effects on the moments of $P(\rho_R)$.
All data are averaged over at least 50 independent runs.

\begin{figure}[t]	
\includegraphics[width=\columnwidth]{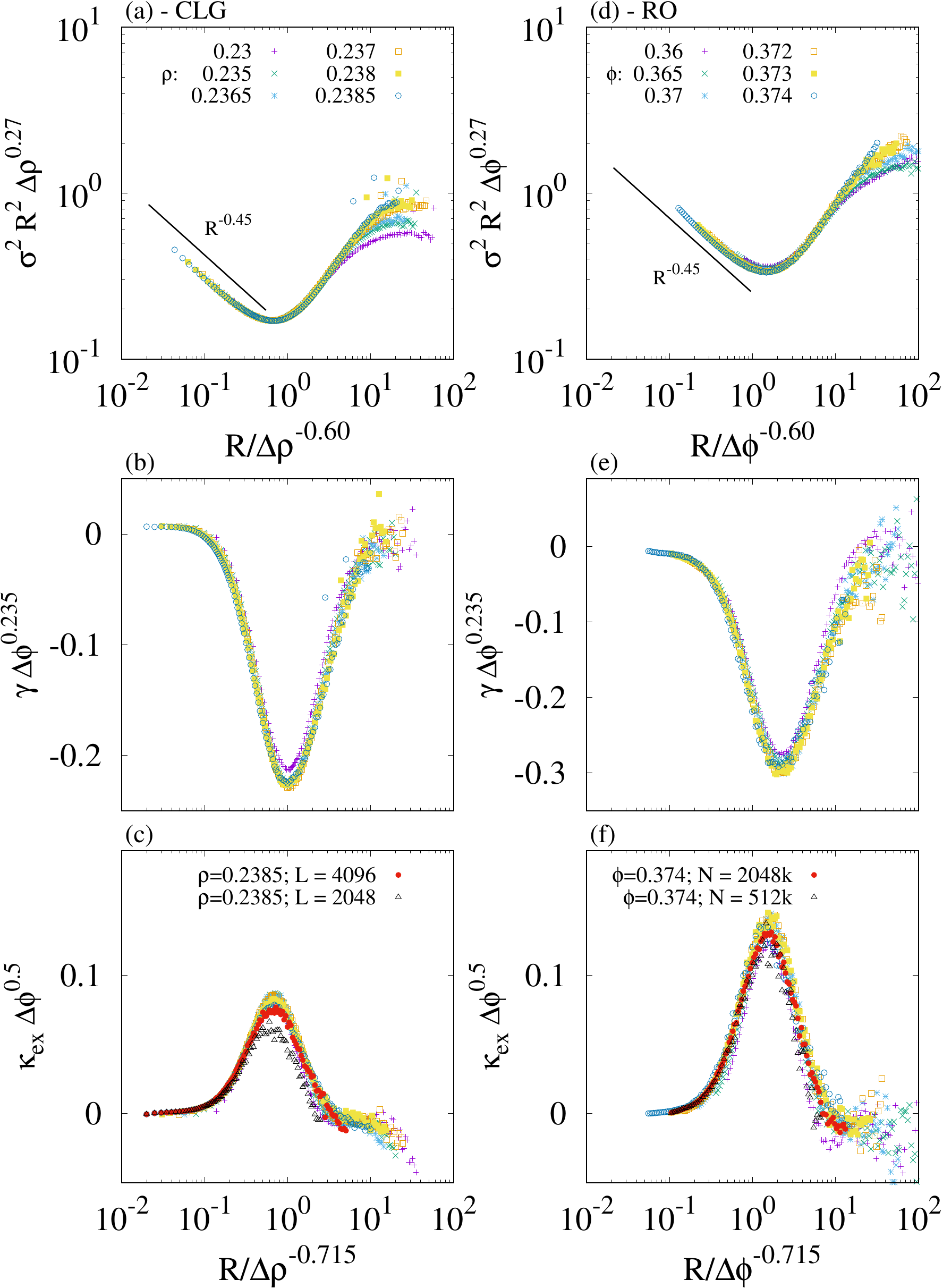}		
\caption{ 
(a,d) Fluctuations $\sigma^2$, (b,e) skewness $\gamma$, and (c,f) excess kurtosis $\ke$ of the coarse-grained density distribution $P(\rho_R)$, as a function of the coarse grained length scale $R$. Panels (a-c) and (d-f) show results for the CLG and the RO model respectively for different densities and volume fractions.
For the CLG model, data consists of systems with linear size $L=8192$, while for the RO model, number of particles $N=8192\times10^3$ are considered. 
Panels (c) and (d) investigate size-effects for the largest densities considered in the respective models. Reducing the linear size by a factor $2$ does not affect our results, while a factor of $4$ leads to considerable size-effects.
\label{fig:scale_collapse}
}
\end{figure}

In the absorbing phase, we observe a suppression of the density fluctuations, with $\sigma^2(R) \sim R^{-\lambda}$ and $\lambda = 2.45$ found to be universal across the broader class of two-dimensional random organization models~\cite{HexnerLevine2015}.
This suppression extends up to the hyperuniformity length scale $\xis\propto  \Delta \rho^{-\nus}$ diverging at the critical point,  where the scaled fluctuations $R^2\sigma^2(R)$ reach their minimum value. 
Departing from previous suggestions that $\nus \simeq \nu_\perp \simeq 0.8$ ~\cite{HexnerLevine2015}, we find instead that $\nus \simeq 0.6$, which in combination with $\lambda \simeq 2.45$,  allows for a collapse of the scaled density fluctuation up to the hyperuniformity length scale, as we show in 
Fig.~\ref{fig:scale_collapse}(a)-(b).
We detail the evaluation of $\nus$ and of other critical exponents, alongside corresponding information on the pair correlation and structure factor, in the Supplementary Material (SM)~\cite{SM}.

\begin{figure}[tb]
\includegraphics[width=\columnwidth]{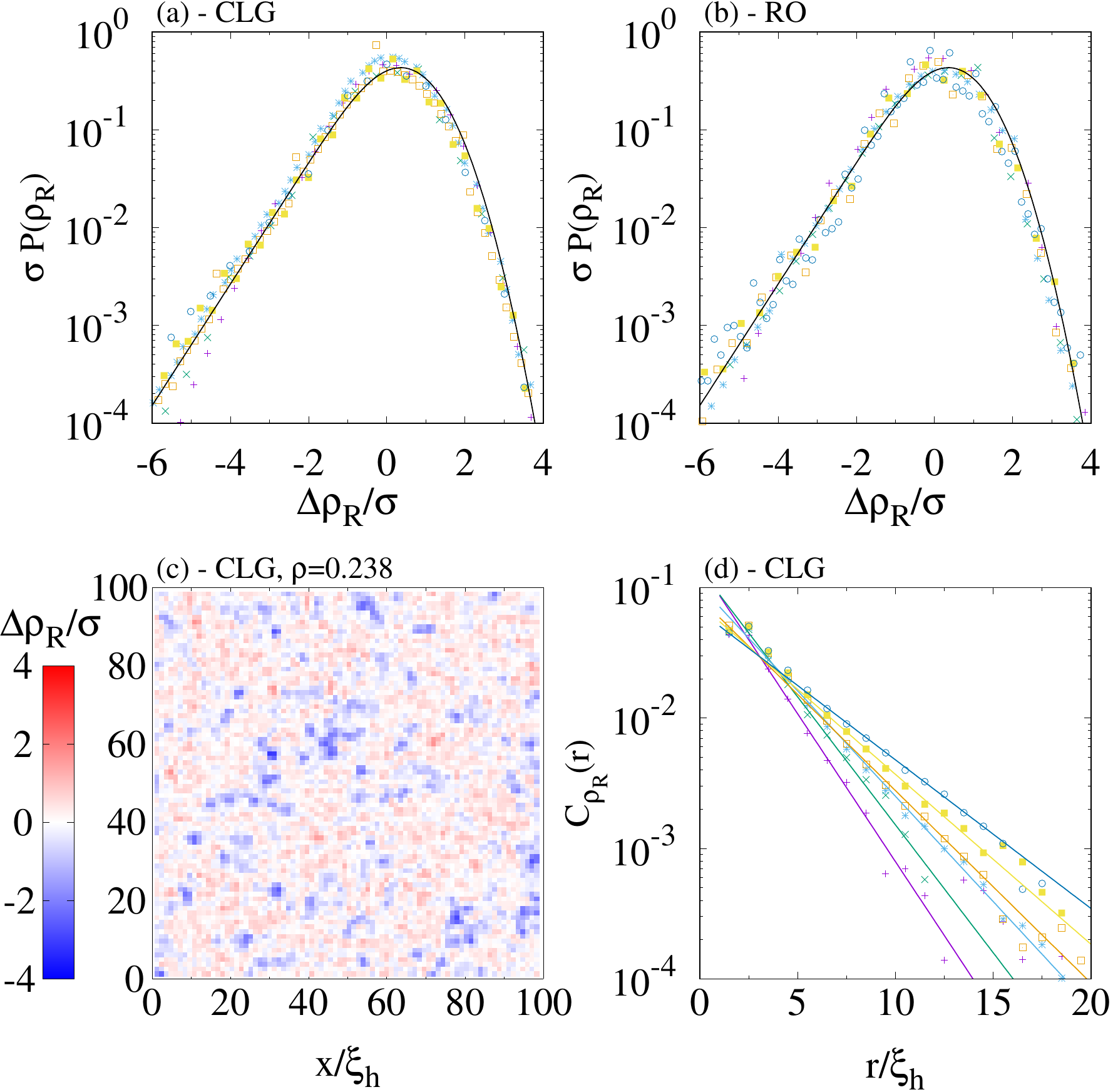}
\caption{
Scaling of the probability distribution of the coarse grained density $\rho_{R = \xis}$,
for (a) CLG  and (b) RO models, with
$\Delta \rho_R = \rho_R-\rho$. 
The black line represents Eq.~\ref{eq:prhoxi} with $A = 0.4$ and $B = 0.7$, in both panels.
(c) Real space map of $\rho_R$ with $R = \xis \simeq 44$ for the CLG model at $\rho = 0.238$. (d) The correlation function of $\rho_{R=\xis}$ decays exponentially, after an initial precipitous drop that occurs at short distances $R \leq \xi$, .
We extract the decay length $\xi_{\rho_R}$ via exponential fits, limited to $R < 15\xis$, represented by the solid lines. 
In (a),(b) and (d), symbols indicate the various densities $\rho$ as labelled in
Fig.~\ref{fig:scale_collapse}.
\label{fig:distribution}
}
\end{figure}

Now, the non-monotonic behavior of the scaled density fluctuations, which themselves do not scale collapse for $R > \xis$, is opposed to the behavior observed in fluctuation suppressed states in models of cell tissues~\cite{ZhengPicaCiamarra2020}, and this suggests the presence of additional correlations in the density field at length scales larger that $\xi_h$.
Therefore, we expect $P(\rho_R)$ to not be Gaussian at fluctuation suppressed length scales, but to become Gaussian at much larger lengths. 
Indeed, we observe in Fig.~\ref{fig:distribution} a non-Gaussian profile for $P(\rho_R)$ at the hyperuniform length scale ($R=\xis$), where a distinctive exponential tail in the low density regime persists that can be approximately described by a universal function, 
\begin{equation}
P(\rho_R) \propto \exp\left(A\frac{\rho_R-\rho}{\sigma}\right) {\rm erfc}\left(B\frac{\rho_R-\rho}{\sigma}\right),
\label{eq:prhoxi}
\end{equation}
where $A$ and $B$ are universal, model independent parameters. 

To investigate the approach to the Gaussian limit, we focus on the $R$ dependence of the skewness $\gamma \equiv \langle [(\rho_R - \rho)/\sigma]^3  \rangle$ and excess kurtosis $\ke \equiv \langle [(\rho_R - \rho)/\sigma]^4  \rangle -3$.
We observe that $\gamma$ and $\ke$ have a non-monotonic dependence on the coarse grained length scale, and exhibit respectively, a minimum and maximum at a length diverging with exponent $\simeq 0.715$~\cite{SM}. 
The extreme values of $\gamma$ and $\ke$ diverge on approaching the transition, with model independent exponents, as illustrated by the scale collapse of Fig.~\ref{fig:scale_collapse}.

To demonstrate that these results are not affected by size-effects, we compare, for the CLG model, data for $L=8192$ with results for $L=4096$ and $L=2048$ - at the highest of densities considered, focusing on the kurtosis and regimes where finite size effects are maximal. 
Fig.~\ref{fig:scale_collapse}c illustrates the good agreement between $L=8192$ and $L=4096$, indicating that our results for $L=8192$ are not affected by size-effects.
Conversely, significant deviations are apparent for $L=2048$. 
The different $\nus$ reported in Ref.~\cite{HexnerLevine2015}, which derived results from $L=1000$, may thus be due to finite size effects. 
Similarly, for the RO model, we find an analogous scenario where results from $N=8192\times10^3$ agree with what is obtained from $N=2048\times10^3$ - at the largest considered value of $\phi$ (Fig.~\ref{fig:scale_collapse}d). 
Hence, size effects only become apparent in systems significantly smaller than the respective largest system sizes of both models that are considered in this Letter.

These results indicate that a sufficiently large system ($L\gg\xis$) can be seen as a tessellation of domains of linear size $\xis$ and density $\rho_R$ distributed in accordance to Eq.~\ref{eq:prhoxi}, within which density fluctuations are suppressed.
Furthermore, the behavior of $P(\rho_R)$ under coarse graining indicates that the coarse-grained density of these domains is spatially correlated, suggesting a hierarchy of self-organization present at different scales.
To visualize this, we illustrate in Fig~\ref{fig:distribution}c a map of the coarse grained density distribution in a spatial region that spans $100\xis\simeq 4200$ for a system of size $L = 8192$ at $\rho = 0.238$ of the CLG model. The observed patches confirms the existence of correlations beyond $\xis$.
We also note that in Fig.~\ref{fig:scale_collapse}f size effects are seen at $\rho=0.2385$ in a system with $L = 2048\simeq 34\xis$, indirectly proving the existence of density correlations extending well beyond $\xis$.

\begin{figure}[!t]
	\includegraphics[width=\columnwidth]{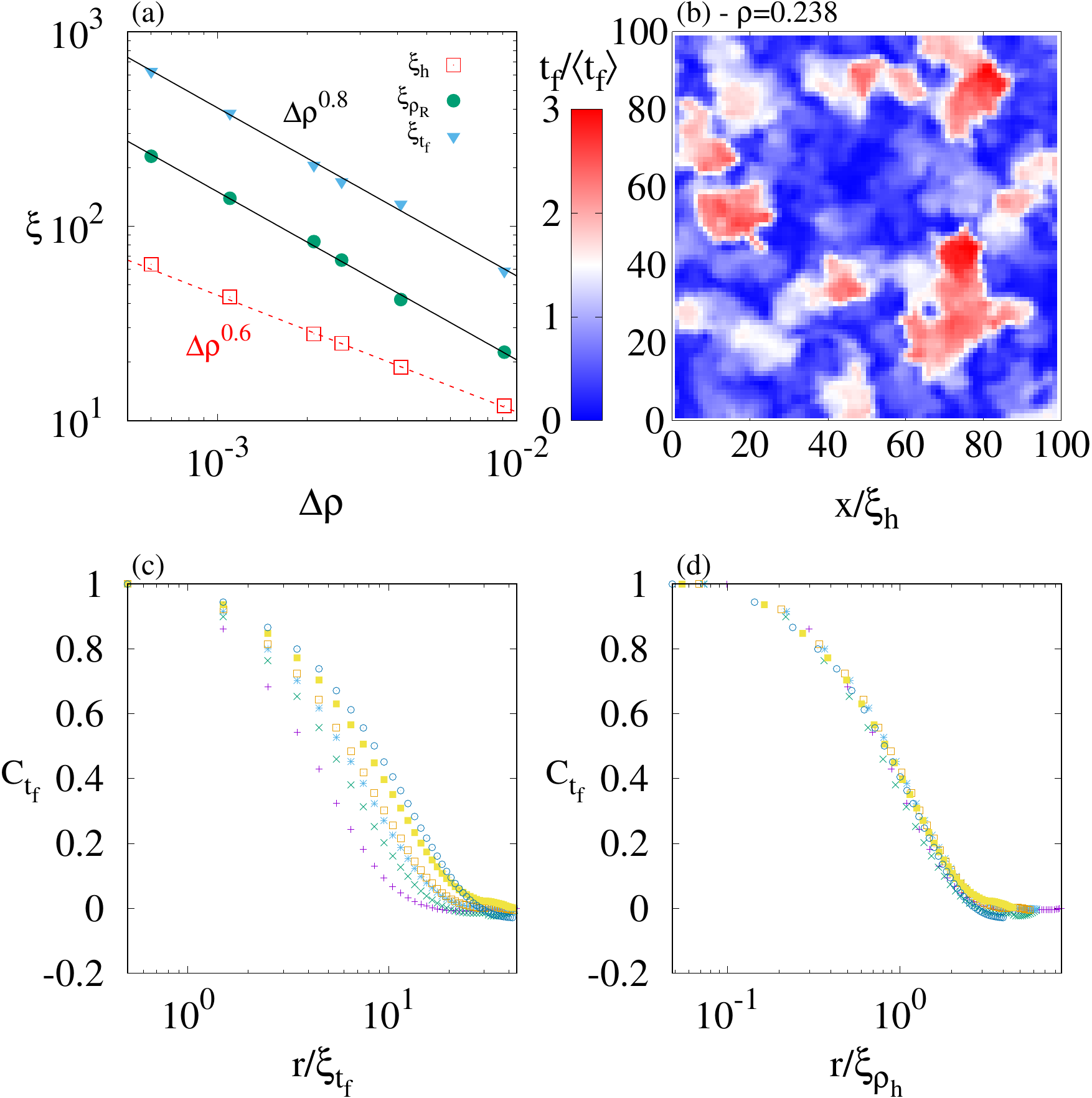}
	\caption{ 
	(a) The hyperuniformity length scale diverges as $\xis \propto \Delta\rho^{-\nus}$, with $\nus \simeq 0.6$, while the correlation length of the coarse grained density (Fig.~\ref{fig:distribution}d)  diverges as $\xi_{\rho_R} \propto \Delta\rho^{-\nu_{\rho_R}}$, with $\nu_{\rho_R} =0.8$.
	The freezing time is correlated over a length scale $\xi_{t_f} \propto \xi_{\rho_R}$.
	(b) Map of the freezing time coarse-grained at length scale $\xis$, for $\rho = 0.238$. 
	(c) and (d) show the correlation function of the coarse-grained freezing time plotted against $r/\xis$ and  $r/\xi_{t_f}$, respectively, where $C(\xi_{t_f}) = e^{-1}$. 
	In (c) and (d), symbols represent various densities labelled in Fig.~\ref{fig:scale_collapse} of the CLG model.
	}
\label{fig:lengthscales}
\end{figure} 

To quantify these additional correlations, we investigate the correlation function of the coarse-grained density
$
C_{\rho_R}(r) \propto \langle \rho_{R}(r) \rho_{R}(0) \rangle - \langle \rho_{R} \rangle^2,
$
at $R = \xis$. 
We discuss in Fig.~S3 in the SM~\cite{SM} the dependence of $C_{\rho_R}(r)$ on $R$.
Fig.~\ref{fig:distribution}d shows that, for the CLG model, this correlation function decays exponentially, $C_{\rho_R}(r) \propto \exp(r/\xi_{\rho_R})$, after a sharp decline at short length scales.
Data corresponding to different densities do not collapse when plotted against $r/\xis$, proving that these observed spatial correlations are a complimentary indication of a hidden additional correlation beyond hyperuniformity.
Indeed, we observe in Fig.~\ref{fig:lengthscales} that, while $\xis$ diverges with the exponent $\nus\simeq 0.6$, $\xi_{\rho_R}$ diverges with the exponent $\nu_{\rho_R} \simeq 0.8$.
We stress that the radial correlation function, which we illustrate in Fig.~S5 in the SM~\cite{SM}, does not reveal these additional correlations, which implies that they are therefore many-body in nature.
These correlations are instead apparent in the two-body correlation function of the coarse grained quantity $C_{\rho_R}(r)$ as they are encoded in $\rho_R$, the probability of finding a region of linear size $R$ that contains $N$ particles.

Noticing that $\nu_{\rho_R} \simeq \nu_\perp$, we investigate the physical origin of this novel length scale focusing on the dynamical process leading to the absorbing states.
Specifically, we study the coarse grained freezing time $t_f$~\cite{cgprocedure}, which is defined as the final time at which a particle transitions from an active to passive state.
Maps of the coarse grained freezing time, as illustrated in Fig.~\ref{fig:lengthscales}b for the CLG model at $\rho = 0.238$, reveal the presence of extended correlations, which we quantify by investigating the correlation function 
$C_{t_f}(r) \propto \langle t_f(0) t_f(r) \rangle-\langle t_f \rangle^2$.
Fig.~\ref{fig:lengthscales}c shows that the $C_{t_f}(r)$ of different densities do not collapse when plotted against $r/\xis$, indicating that these correlations are also not set by the hyperuniformity length scale.
Instead, we collapse the correlation function in Fig.~\ref{fig:lengthscales}d by plotting them against $r/\xi_{t_f}$, where the freezing time correlation length scale is defined by $C_{t_f}(\xi_{t_f}) = e^{-1}$.
The length scale $\xi_{t_f}$ diverges on approaching the absorbing phase transition as $\Delta \rho^{-0.8}$, as we demonstrate in Fig.~\ref{fig:lengthscales}a, and is therefore proportional to $\xi_{\rho_R}$.
The additional spatial correlations in the density field beyond hyperuniformity, therefore, reflect the spatial correlation in the dynamics leading to the absorbing state.

\begin{figure}[!t]
\includegraphics[width=\columnwidth]{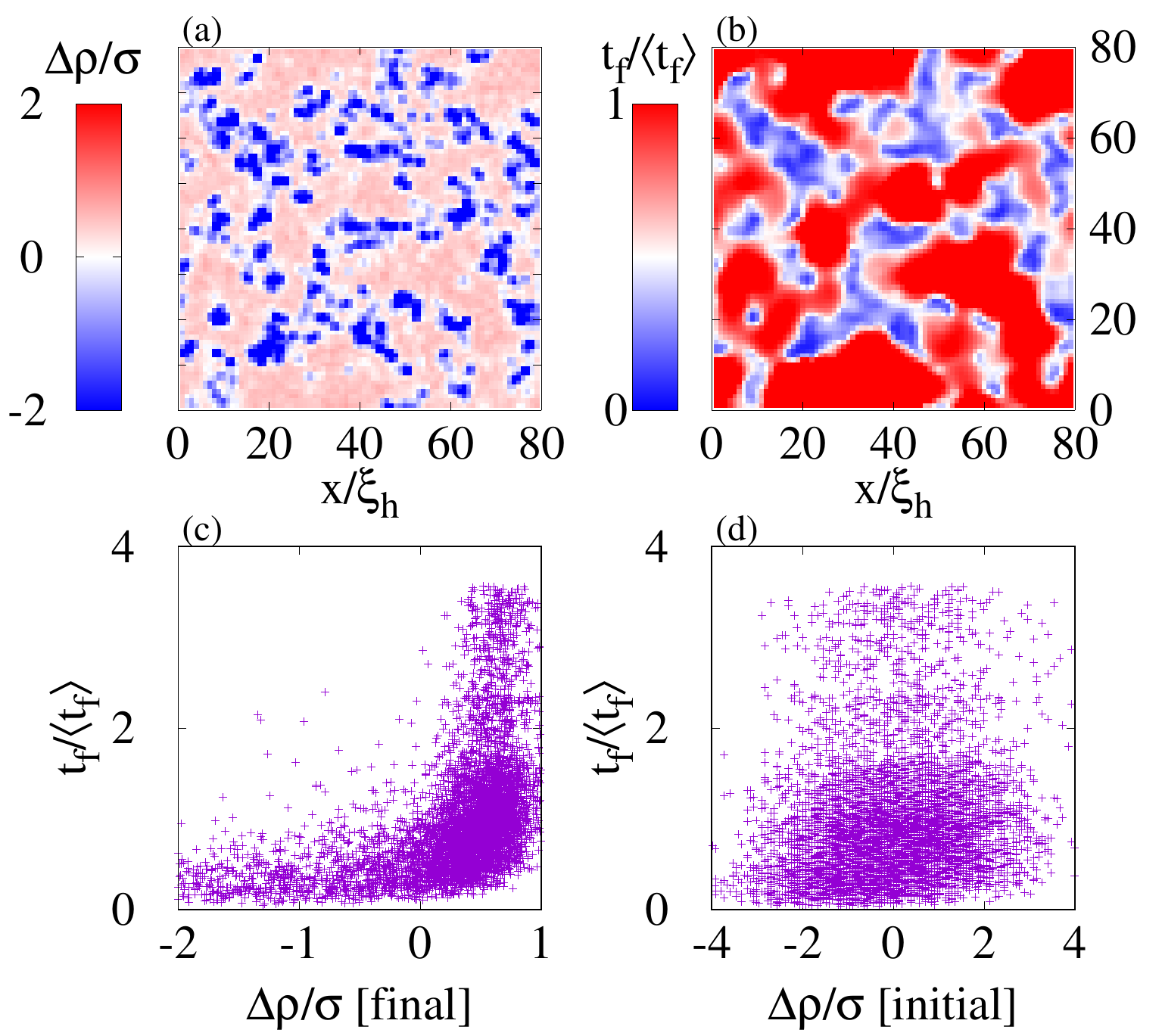}
\caption{ 
Maps of the (a) density and (b) freezing time coarse grained over length scale $\xi_{h}$ and averaged over $100$ realizations that share the same initial configuration.
The visual similarity between panel (a) and (b) is quantified in (c) through the scatter plot of the average coarse-grained freezing time versus the average coarse-grained density of the final configuration.
(d) The freezing time is not correlated to fluctuations of the density of the initial configuration. 
All data refer to the CLG model at $\rho = 0.238$ and $L=4096$. 
}
\label{fig:propensity}
\end{figure}

These results prove that a single length scale controls the correlation of the coarse-grained density and that of the freezing time.
The correlation between these two quantities, however, is not immediately apparent by a comparison of the coarse-grained density of the final configuration of Fig.~\ref{fig:distribution}c and that of the coarse-grained freezing time, Fig.~\ref{fig:lengthscales}b.
To unveil these correlations we compare the two maps averaged over 100 runs~\cite{Harrowell2004} sharing the same initial configurations.
These so-called isoconfigurational averages mitigate the effect of dynamical noise.
The resulting maps, which we show in Figs.~\ref{fig:propensity}a and b, are clearly similar, as regions with a low average coarse-grained density correspond to regions with a low average coarse-grained freezing time.
Plotting these quantities against one another in a scatter diagram as in Fig.\ref{fig:propensity}c makes their correlation apparent. Importantly, we remark that these correlations do not originate from fluctuations of the coarse-grained density in the initial configuration.
Indeed, we show in Fig.~\ref{fig:propensity}d that the average coarse-grained freezing time does not correlate with the fluctuations of the coarse-grained density of the initial configuration.

These observations establish an intriguing analogy between the behaviour of the coarse-grained density distribution $P(\rho_R)$ on increasing $R$, and that of the displacement of the particles of supercooled liquids $P(\Delta r_t)$ on increasing the observation time $t$. 
The fluctuations of $\rho_R$ are suppressed up to a length scale $\xis$ that diverges at the active-absorbing phase transition, and the convergence to CLT is recovered gradually at larger length scales; Similarly, the fluctuations of $\Delta r_t$ are suppressed up to the relaxation time $\tau$ that ideally diverges at the glass transition, the predictions of CLT being progressively recovered at longer timescales.
In the CLG and RO models, $\ke$ peaks at lengths that scale diverging at the transition, where the peak height also diverges; Likewise, in supercooled liquids, $\ke$ peaks at a time scaling with $\tau$, and its peak value diverges at the transition~\cite{Xu2018}.
Furthermore, the heterogeneites of the coarse grained density at $R = \xis$ observed in RO models, Fig~\ref{fig:distribution}c, reflect that of the displacements of the particles at the relaxation time in supercooled liquids ~\cite{Berthier2011a}. 
Finally, in both cases, the correlations between these spatial heterogeneities and the initial configuration of the system are revealed via isoconfigurational averages~\cite{Harrowell2004}.
The systems only differ in their skewness where $\gamma = 0$ in liquids as a consequence of the invariance of the equation of motion under time reversal.

In this Letter, we established the existence of hidden many-body correlations, beyond what is captured by the suppression of density fluctuations, in two-dimensional models exhibiting an active-absorbing phase transition. 
Qualitatively, the scenario here discussed appears to also hold in three dimensions~\cite{SM}. 
Specifically,
density fluctuations are suppressed up to a length scale that diverges at the transition with the exponent $\nus \simeq 0.6$ while conversely, many-body density correlations extend up to a longer length scale that diverges with the exponent $\nu_\perp \simeq 0.80$.
The presence of similar findings beyond the current context is an exciting avenue that demands further investigation. 
These explorations may, in turn, provide means for further taxonomy and classification of fluctuation suppressed or hyperuniform disordered systems.
Specifically, we envisage a parallel with the taxonomy recently introduced for diffusive systems~\cite{Wang2012, Chechkin2017}, where four main classes are identified based on the behaviour of the second  (Fickian/non-Fickian) and higher moments (Gaussian/non-Gaussian) of the displacement field probability distribution.
From this perspective, the CLG and RO models are considered hyperunifiorm but non-Gaussian systems, random jammed sphere packings~\cite{Klatt2016} are (effectively) hyperuniform and Gaussian, while the Voronoi model for cell tissue~\cite{ZhengPicaCiamarra2020} and Quantizer problems ~\cite{KlattTorquato2019} are (effectively) hyperuniform and Gaussian.
Seemingly, non-Gaussian behavior appears to occur in systems exhibiting an absorbing transition while Gaussian behavior exists in jammed solids. 
Further work in this direction is certainly needed. 
Future investigations may also consider the possibility of artificially tuning the Gaussian behavior through local particle displacements~\cite{Klatt2020}.

More generally, these analyses based on the higher moments may provide for additional tools in probing the possible causal role of hyperuniformity in the self-organization of disordered systems by further characterizing the approach to criticality of nonequilibrium phase transitions.

\begin{acknowledgments}
We thank S. Torquato for  insightful comments, and acknowledge support from the Singapore Ministry of Education through the Academic Research Fund MOE2017-T2-1-066 (S) and
MOE2019-T1-001-03 (S).
\end{acknowledgments}


\pagebreak
~\newpage
\onecolumngrid

In this supplementary, we provide additional figures that compliment various results contained in the main text. A summary of figures included are as follows:

\section{Determination of the critical exponents}
We detail here our evaluation of the exponent $\nu_h$ that controls the divergence of the hyperuniform length scale at the critical point from the absorbing phase, $\xis \propto (\rho_c-\rho)^{-\nu_h}$.
We have estimated $\nu_h \simeq 0.60$, a value sensibly different from what has been previously reported in \cite{HexnerLevine2015}, $\nu_h \simeq 0.80$.

\begin{figure}[!b]	
		\includegraphics[width=0.6\columnwidth]{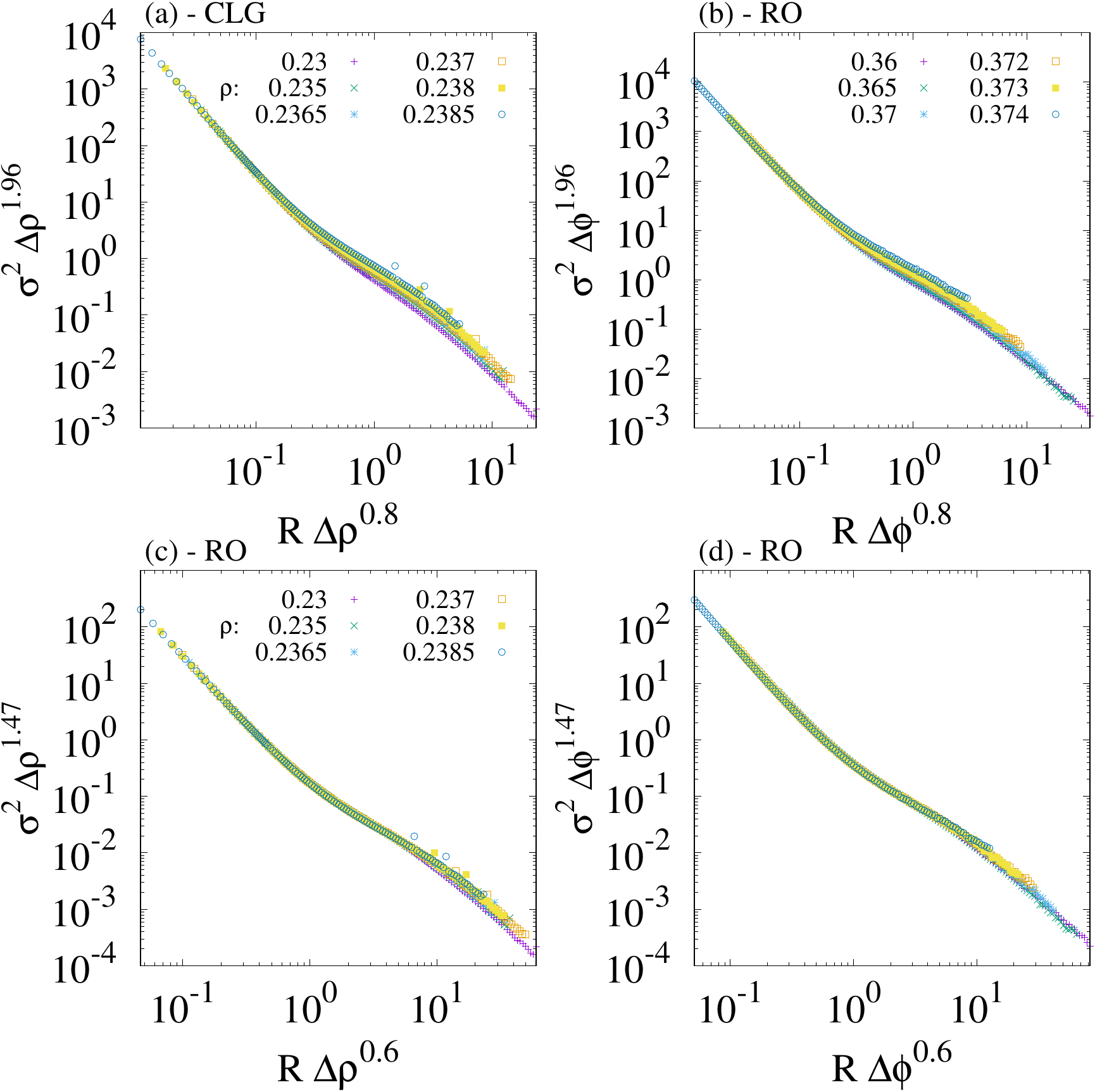}		
	\caption{
	Scaling of the fluctuations of the coarse-grained density, for the CLG model in (a) and (c), and for the RO model in (b) and (d). In (a) and (b), we fix $\nu_h = 0.8$, while in $(c)$ and $(d)$ we fix $\nu_h = 0.6$.
	}
	\label{fig:0}
\end{figure}

To estimate the scaling exponents, previous works have considered that since $\sigma^2(R) \propto R^{-\lambda}$ with $\lambda \simeq 2.45$, in the fluctuation suppressed region, then  $\sigma^2(R) \Delta\rho^{-\nu_h\lambda}\propto (R/\Delta\rho^{-\nu_h})^{-\lambda}$.
Therefore, the value of $\nu_h$ can be estimated via a data collapse.
In Fig.~\ref{fig:0}(a) and (b) we show that a reasonable data collapse is obtained for $\nu_h \simeq 0.8$ in the fluctuation suppressed region, for both models. 
Panel (c) and (d), however, demonstrate that $\nu_h \simeq 0.6$ provides a much better data collapse.
This suggests that the exponent $0.8$ estimated in previous works might not be accurate especially considering the small system sizes considered there.

\begin{figure}[!t]	
	\includegraphics[width=0.5\columnwidth]{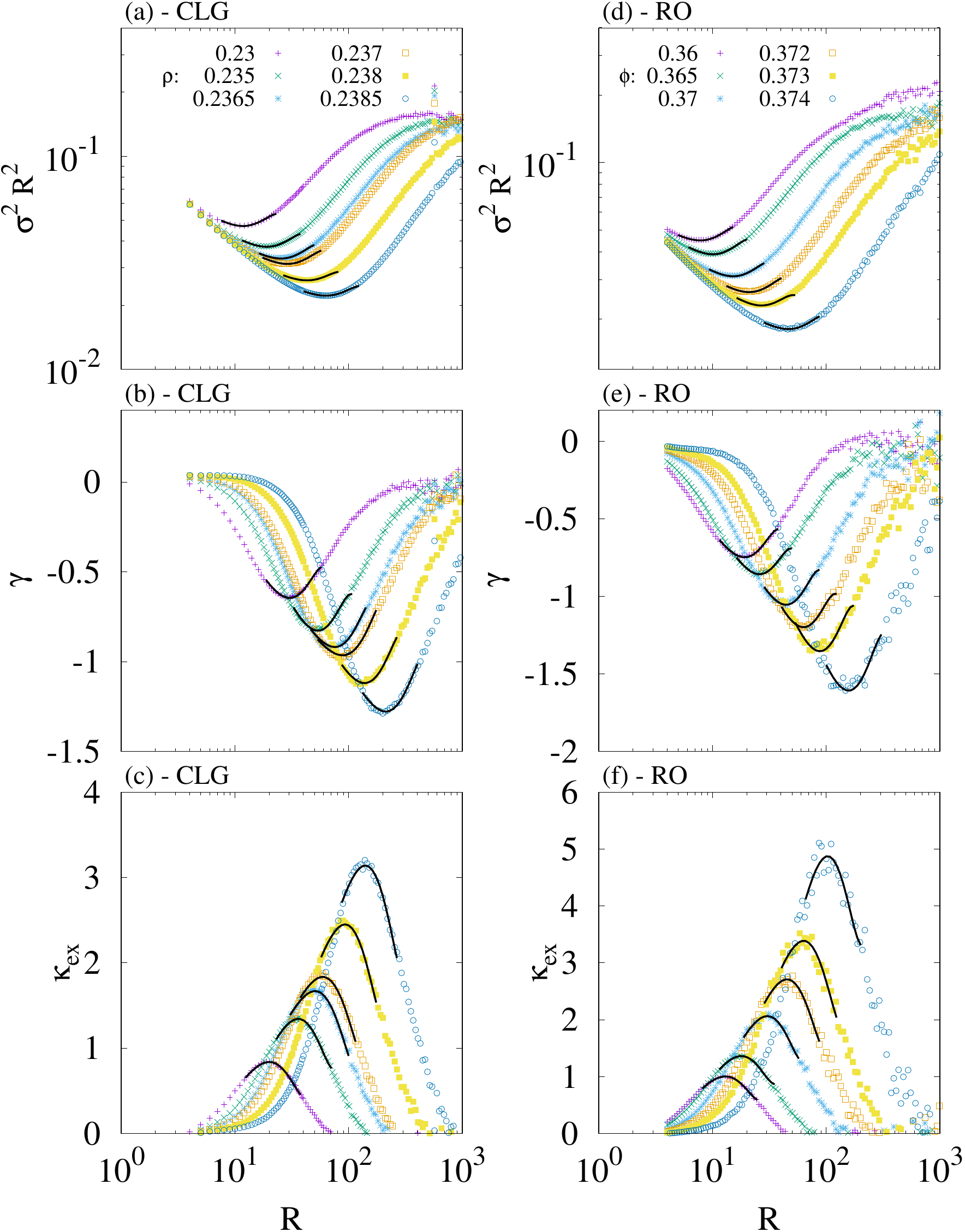}		
	\caption{
	Dependence of the scaled fluctuations $\sigma^2 R^2$, skewness $\gamma$ and kurtosis $\kappa$ of the probability distribution for the coarse grained density in the CLG and RO models.
	We estimate the length scales $\xis$, $\xi_\gamma$ and $\xi_\kappa$ at which these quantities attain their extreme values via polynomial fits, shows as full lines.
	}
	\label{fig:1}
\end{figure}

To quantitatively evaluate the critical exponents without resorting to the visual impressions provided by a data collapse, we investigate the coarse-grained dependence of the scaled fluctuations, $R^2\sigma^2(R)$, as well as the skewness, $\gamma$, and the kurtosis, $\kappa$. 
It is instructive to focus on $R^2\sigma^2(R)$ as this quantity exhibits a minimum at the $R$ value at which the density fluctuations are maximally suppressed, signalling the crossover from the suppressed fluctuation region, $\sigma^2(R) \propto R^{-\lambda}$, to the central limit theorem scaling, $\sigma^2(R) \propto R^{-2}$.
The non-monotononic dependence on $R$ of these quantities, illustrated in Fig.~\ref{fig:1}, allows the unambiguous identification of the $R$ values corresponding to their extremes, which we indicate as $\xi_x$ where $x = \sigma^2,\gamma$ and $\kappa$. 
We evaluate $\xi_x$ via polynomial fits close to their extreme values, also illustrated in the figure.

\begin{figure}[!h]	
	\includegraphics[width=0.5\columnwidth]{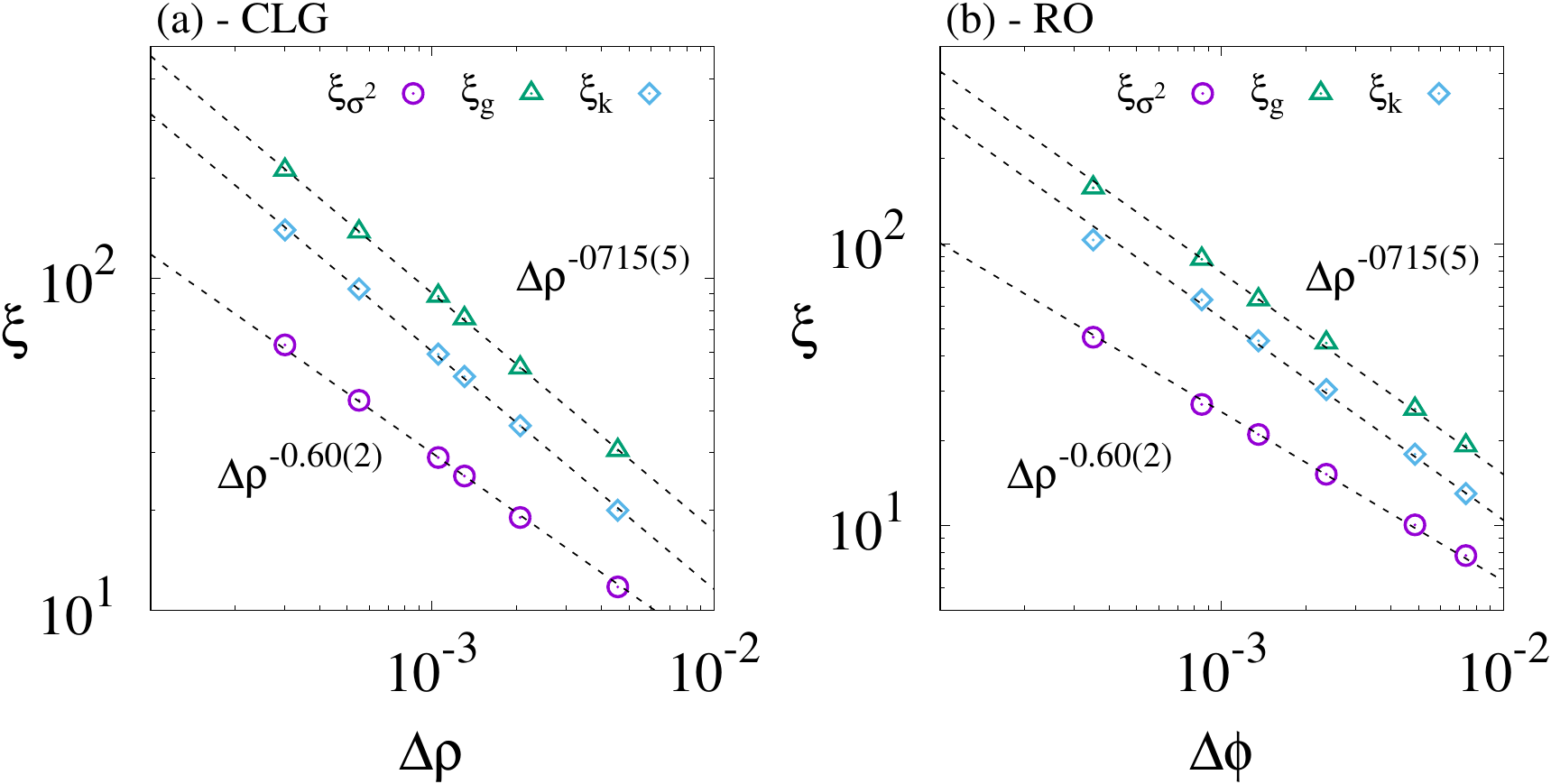}		
	\caption{
	length scales obtained from the extreme values of the scaled fluctuations $\sigma^2 R^2$, skewness, $\gamma$, and kurtosis. Power law fits to these length scales are indicated by the black dashed lines. 
	}
	\label{fig:2}
\end{figure}

Fig.~\ref{fig:2} shows that both for the RO and the CLG models, the length scale $\xis$ diverges at the critical point with exponent $\nu_h \simeq 0.6$. 
Conversely, the length scales $\xi_\gamma$ and $\xi_\kappa$ diverge with critical exponent $\simeq 0.715$.  

\section{Correlation of the coarse-grained density}
\begin{figure}[!h]	
	\includegraphics[width=0.8\columnwidth]{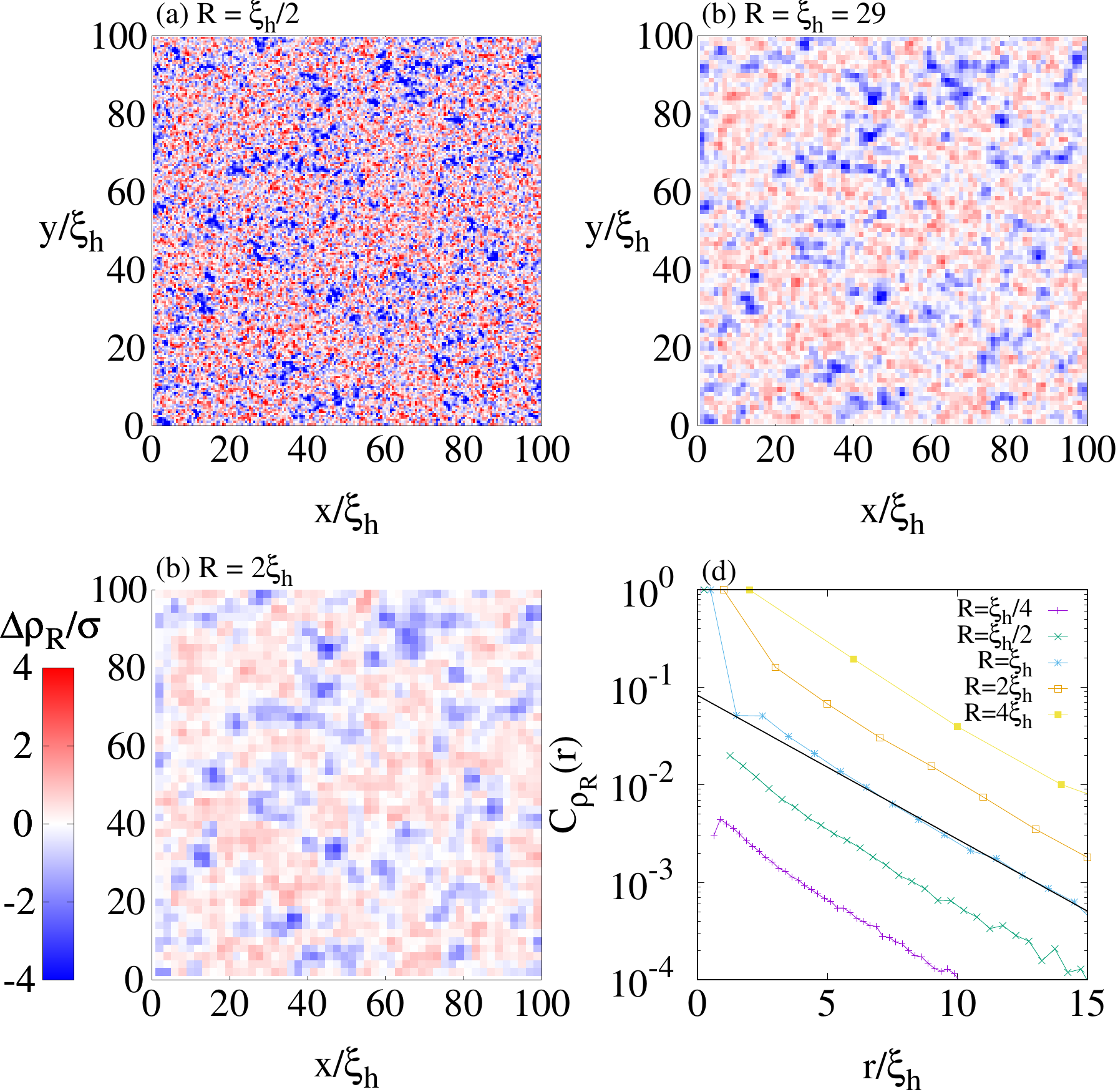}	
	\caption{
	Maps of the density field coarse-grained at length scale $R = \xis/2$ (a), $R = \xis$ (b) and $R = 2\xis$ (c), for the CLG model at $\rho = 0.237$, where $\xis \simeq 29$ lattice sites. Panel (d) shows that a same length scale controls the decay of correlation function of the coarse-grained density, at large distances. 
	}
	\label{fig:3}
\end{figure}

In Fig. 2 of the main text we have extracted a length scale from the decay of the correlation function of the density coarse-grained over a length scale $R = \xis$. 
In Fig.~\ref{fig:3}(a)-(c) we compare maps of the density coarse-grained at different length scales. Panel (d) shows that the decay length does not critically depend on $R$. The amplitude of the correlation, however, does depend on $R$, so that the amplitude vanishes as $R$ decreases. 

\begin{figure}[!h]	
	\includegraphics[width=0.8\columnwidth]{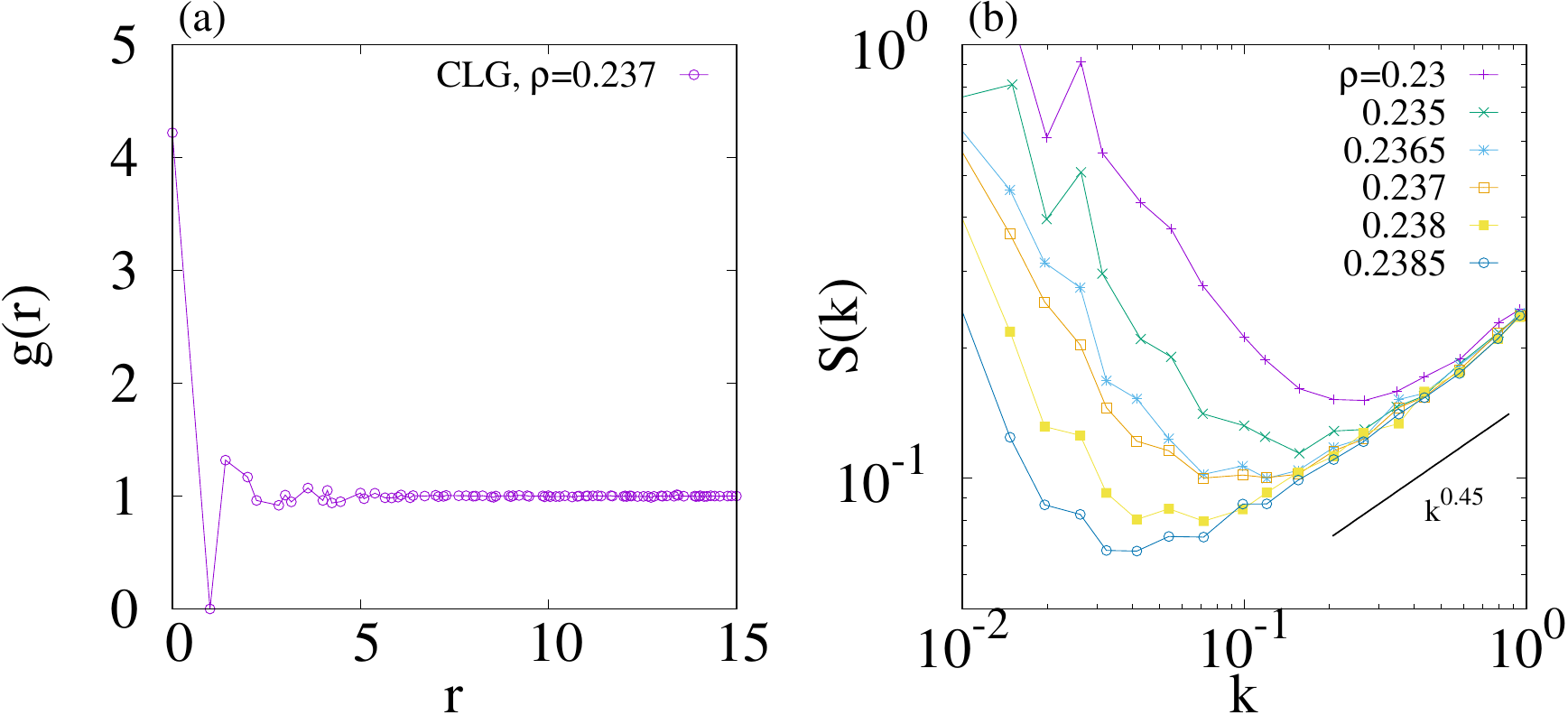}
	\caption{
	(a) The Radial correlation function of the CLG model, here shown at $\rho = 0.237$, attains the ideal gas limit on a length scale much shorted than the hyperuniform length, which at this density is $\xis \simeq 29$.
	(b) structure factor at different densities. 
	}
	\label{fig:4}
\end{figure}

We stress that the correlation length of the coarse-grained density does not reflect correlations in the radial distribution function.
Indeed, the radial correlation  function attains its limiting value on a length scale much smaller than $\xis$, and hence of $\xi_{\sigma^2}$, as apparent from Fig.~\ref{fig:4}a.
Similarly, the structure factor, illustrated in Fig.~\ref{fig:4}b, suggests suppression of the density fluctuations up to a length scales increasing on approaching the critical point, and not the existence of other characteristic length scales. We note that these results for the  structure factor are analogous to those previously observed in \cite{HexnerLevine2015}.

\newpage
\section{CLG model in thee spatial dimensions}
\begin{figure}[!h]	
	\includegraphics[width=0.9\columnwidth]{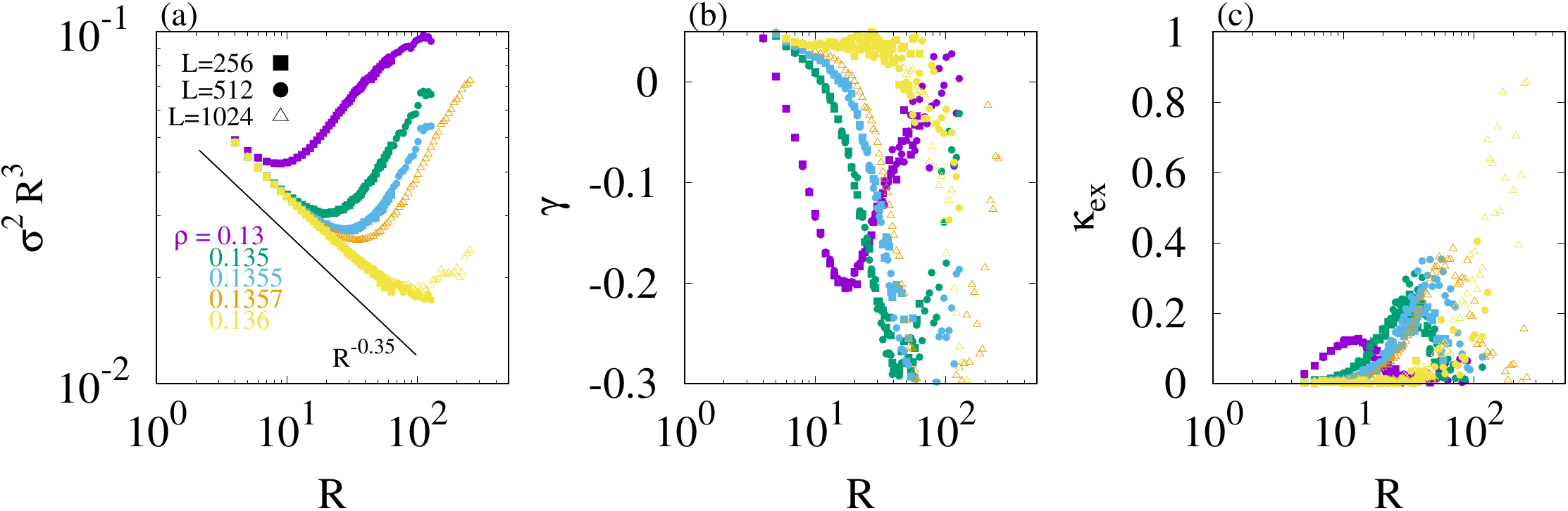}
	\caption{
	(a) scaled density fluctuations, (b) skewness and (c) excess kurtosis of the distribution of the density coarse-grained at a length scale $R$, for the CLG model in three spatial dimensions. 
	Symbols and colour identify the system sizes and densities, as in the legend in (a).
	}
	\label{fig:5}
\end{figure}

To check the dimensionality dependence of our findings, we have performed simulations of the CLG model in three spatial dimensions.
Fig.~\ref{fig:5} illustrate the dependence of (a) scaled variance, (b) skewness and (c) excess kurtosis on the coarse-graining length scale.
Results are qualitatively analogous to those observed in two dimensions, e.g. Fig.~\ref{fig:2}(a), (b) and (c), suggesting that the general phenomenology remains the same.
We are unable to perform a detailed investigation of the three dimensional results due their huge computational cost. 
The noisy data in Fig.~\ref{fig:5}, in fact already involve simulations of systems with 100 millions of particles.

\newpage
\section{Other models}
\begin{figure}[h]
	\includegraphics[width=0.55\columnwidth]{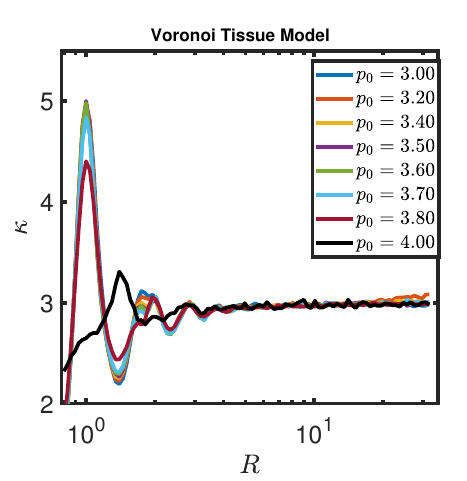}
	\caption{Monotonic and $p_0$ independent decay of the kurtosis to Gaussian values at very short length scales in the solid phase of the voronoi tissue model which is known to be effectively hyperuniform at $p_0^{\star}\simeq 3.81$ and above.}
	\label{fig:s7}
\end{figure}

We illustrate in Fig. \ref{fig:s7} the dependence of the Kurtosis on the coarse graining length scale for the Voronoi model of cell tissue.
In this model, the parameter $p_0$ drives a transition/crossover from a solid to a fluid phase at $p_0^* \simeq 3.81$, in the direction of increasing $p_0$.
The system is known to be effectively hyperuniform at the critical point and in the fluid phase~\cite{ZhengPicaCiamarra2020}. 
In the figure, we observe a rapid monotonic decay of the kurtosis to its Gaussian value independent of its distance from the critical point.
The kurtosis does not indicate the existence of additional correlations beyond hyperuniformity. 
The Voronoi model, therefore, is an example of a effectively hyperuniform and Gaussian systems.

\begin{thebibliography} {200}
\bibitem{cgprocedure}
We evaluate the variable $x$ coarse-grained on a length scale $R$, $x_R$, dividing the system in $(L/R)^d$ non overlapping square windows, and averaging $x$ within each window.
\end{thebibliography}
\end{document}